\documentclass[12pt]{article} 
\usepackage{graphics}
\usepackage{latexsym} 
\newcommand{\be}{\begin{equation}} 
\newcommand{\ee}{\end{equation}}

\newcommand{\ra}{\rightarrow} 
\newcommand{\bea}{\begin{eqnarray}} 
\newcommand{\eea}{\end{eqnarray}}

\newcommand{\nn}{\nonumber}

\newcommand{\we}{\wedge}

\newcommand{\N}{{\cal{N}}} 
\newcommand{\F}{{\cal{F}}}
 
\newcommand{\tilda}{\tilde} 
\begin{document} 

\begin{flushright}
CALT-68-2417\\
UCLA/02/TEP/42\\
hep-th/0212252
\end{flushright}
 
\bigskip \bigskip 
\centerline{\large \bf Exact $\N=2$ Supergravity Solutions With Polarized Branes}
\bigskip 
\bigskip       
\centerline{{\bf Iosif Bena$^1$ and Calin Ciocarlie$^2$}}  
\medskip 
\centerline{$^1$ Department of Physics and Astronomy} 
\centerline{University of California} 
\centerline{Los Angeles, CA  90095}
\medskip 
\centerline{$^2$ Department of Physics} 
\centerline{California Institute of Technology} 
\centerline{Pasadena, CA 91125}

\centerline{{\rm iosif@physics.ucla.edu, calin@theory.caltech.edu} }
\bigskip \bigskip 

\begin{abstract}
We construct several classes of exact supersymmetric supergravity solutions describing  D4 branes polarized into NS5 branes and F-strings polarized into D2 branes. These setups belong to the same universality class as the perturbative solutions used by Polchinski and Strassler to describe the string dual of  $\N=1^*$ theories. The D4-NS5 setup can be interpreted as a string dual to a confining 4+1 dimensional theory with 8 supercharges, whose properties we discuss. By T-duality, our solutions give Type IIB supersymmetric backgrounds with polarized branes. 
\end{abstract}
\newpage

\section{Introduction}

Ever since the remarkable discovery of the AdS-CFT duality \cite{m} there has been a lot of interest in finding supergravity duals to 4 dimensional field theories with reduced supersymmetry, and to use these duals to understand real world phenomena like confinement or the generation of a mass gap.

In several cases the supergravity dual of the field theory is pure geometry \cite{ks,mn,pw,dzf,joe2,lerda,alex-amanda}, and the exact supergravity solution, although challenging, was found. In other cases, like the $\N=1^*$ theory, the string/supergravity dual (found by Polchinski and Strassler in \cite{joe}) contains D3 branes polarized \cite{myers} into 5 branes, and the exact geometry is still not known.

This paper attempts to make one step in that direction. We find exact supergravity solutions with polarized branes and with 8 supercharges. These solutions describe D4 branes polarized into NS5 branes, and F1 strings polarized into D2 branes. They are very similar to the Polchinski-Strassler (PS) case, both because polarization takes place in the near horizon geometry of the branes, and because the fields inducing it are tensor harmonics on the transverse space. 

In fact, by T and S duality, these solutions give Type IIB exact solutions containing D3 branes smeared along one direction, which polarize into cylindrical NS5 or D5 branes. These solutions are dual to a limit of the Coulomb phase of the $\N=4$ Super Yang Mills, which can have screening or confining vacua when the $\N=4$ supersymmetry is broken to $\N=2$. As we will see, the radius and orientation of the cylinders parameterize a moduli space of vacua, for each type of (p,q) 5-brane.


{\bf Outline}

We first perform a perturbative investigation of the polarization of D4 branes into NS5 branes, along the lines of \cite{joe}.  
As explained in \cite{imsy}, supergravity in the near-horizon geometry of D branes describes a certain strongly coupled regime of the field theory living on these branes. Both sides of this duality can be perturbed. Introducing an operator in the Lagrangian of the field theory side is dual to turning on a non-normalizable mode of the corresponding supergravity field in the bulk \cite{bdhm-bklt}. 

In the Polchinski-Strassler case, the  3+1 dimensional $\N=4$ Super Yang Mills theory was perturbed to the $\N=1^*$ theory by giving mass to the 3 chiral multiplets.  This was dual to perturbing the $AdS_5 \times S^5$ geometry with RR and NSNS 3 forms along the space transverse to the branes. These forms were responsible for polarizing the D3 branes in (p,q) 5 branes. The resulting setups were dual to the different phases of the $\N=1^*$ theory, and made visible many features of this theory.

In chapter 2, we similarly perturb the near horizon background of a large number of D4 branes with the operator corresponding to a mass term for the chiral multiplet in the  4+1 dimensional $\N=1$ theory on the branes. This operator preserves 8 of the original 16 supercharges, and transforms in the {\bf 10} of the $SO(5)$ R symmetry group. It corresponds in the supergravity dual to a non-normalizable mode of the RR 2-form and NSNS 3-form field strengths on the 5 dimensional space transverse to the branes. 

We will find that $N$ D4 branes can polarize into $k$ NS5 branes only for a very specific value of transverse field perturbation: $F_2 \sim \frac {k}{Ng_s \sqrt {\alpha'}}$. For all other values no polarization happens. Moreover, our analysis shows that the polarization radius is a modulus. 
An identical phenomenon happens when F1 strings polarize into D2 branes \cite{f1}\footnote{This phenomenon was called in \cite{f1} ``Aut Caesar aut Nihil'', and proved to be the key to unearthing the exact supergravity solutions describing polarized branes.}. 

Since the radius is a modulus, it is natural to suspect that these configurations could descend from a Coulomb branch configuration of M5/M2 branes in M-theory. Moreover, all the fields present could descend from the fields of the M5/M2 brane supergravity solution by a twisted Melvin reduction. It is therefore not hard to see what the full picture is:

If we have for example $N$ M5 branes uniformly spaced on a circle, the angle between two of them is $\Delta \phi=2\pi/N$. If one compactifies with a twist of $2\pi/N$, the upper end of an M5 brane is joined with the lower end of its neighboring M5 brane. Thus, the whole Coulomb branch descends into a configuration of $N$ D4 branes polarized into {\bf one} NS5 brane. If one increases the twist $k$ times, the upper end of an M5 brane is joined with the lower end of its $k$'th neighbor, and this gives $k$ chains of M5 branes, which descend into $N$ D4 branes polarized into $ k$ NS5 branes.

For all values of the twist which do not match an M5 brane end with another, the descending configuration has no type IIA brane interpretation (it would be like $N$ D4 branes polarized into a configuration with a noninteger NS5 brane charge). Therefore, compactifications with twists that do not match the brane ends only give consistent type IIA solutions when all the 5 branes are coincident. 

Since the perturbation fields are proportional to the twist, we can see that the above picture matches perfectly the one obtained via the Polchinski-Strassler analysis. The discrete set of values of the fields for which the D4 branes polarize corresponds to the discrete set of twists compatible with the M5 branes being on the Coulomb branch. 

Moreover, the Killing vectors of the M-theory solution do not depend on the radius. Hence, a twist by $2 k \pi/N$ will match the brane ends at any radius. This implies that the descending configuration will be a solution at any radius, and therefore the polarization radius is a modulus, exactly as the field theory analysis implies.

As a side note, if the Killing vectors had different radial dependence, the twist would match the ends of neighboring branes only at certain values of the radius. This would give the possible radii of polarization of D4 branes into one NS5 brane. At a different radius, the ends of ``next of neighbor'' branes would match, and this would give $N$ D4 branes polarized into two NS5 branes. It is possible that this intuitive picture of matching brane ends could be useful in attempting to find the full solution in situations where the radius is not a modulus.

The immediate bonus of the above picture is finding exact Polchinski-Strassler-like IIA solutions with polarized branes by simply reducing with a twist M-theory supergravity solutions with branes spread on a circle. In chapter 3 we will find these solution, and show that they reduce to the first order solution obtained in chapter 2. We will also link the boundary theory fermion mass parameters to the M-theory twists and show that the supergravity solution preserves 8-supercharges, just as expected from the gauge/gravity analysis.

One can also give an identical description to the polarization of F1 strings into D2 branes described in \cite{f1}. In that case the M2 branes on the Coulomb branch are compactified with a twist which matches their ends. This gives a geometry with F1 strings polarized into D2 branes. The radius is again a modulus, and this is consistent with the Killing vectors for  $x_{11}$ and $\phi$ having no radial dependence. The compactification twist preserves 8 supercharges, and can be again related to the masses of the fermion bilinears turned on on the boundary theory to induce polarization. This exact solution is discussed in chapter 4.

In fact, both the twisted M2 and M5 supergravity backgrounds (without the branes being polarized) have recently been obtained by Figueroa O'Farill and Simon \cite{fs}. These solutions are basically superpositions of the supersymmetric flux 5 brane with D4 branes and F1 strings respectively.
The new feature of our supergravity solutions is that for certain values of the fluxes, the D4 branes/F1 strings can polarize into NS5/D2 branes, and that moreover, the polarization radius is a modulus.
Thus, the most general $\N=2$ exact solution we can write contains several D4-NS5 (or F1-D2) concentric circles of different radii, and different orientations.
One can also generate F1-D2 solutions with $\N=1$ supersymmetry, which can have 2 different kinds of F1-D2 solitons, at various radii and orientations. 

Using our methods it is also possible to obtain nonsupersymmetric exact solutions with polarized branes \footnote{Such nonsupersymmetric solutions have been obtained in the past via Melvin reductions \cite{non}.} Indeed, as long as the twist along the circle where the branes are spread matches their ends, one can twist along other directions by arbitrary amounts, and still obtain a good solution. Supersymmetry was necessary in PS-like setups to control the backreaction of the various fields on the metric. However, here we have the {\bf exact} metric, with the polarized branes, and we know that our setup is a solution simply because it is the compactification of an M-theory solution along a Killing direction. 

One of the hopes of the authors is that these exact solutions could be used to find the full geometry of the Polchinski-Strassler setup. Indeed, the equations for the fields have almost identical form, and the interplay of the different RR and NSNS fields is similar. One possible route would be to use a similar ansatz and to wrestle it through the full type IIB equations of motion. Another possible route would involve using Kaluza-Klein twisted reductions of similar spirit to ours, and combining them with dualities to try to obtain the metric without going through the IIB equations.   

In chapter 5 we use T-duality to obtain exact Type IIB supergravity backgrounds containing D3 branes polarized into cylindrical (p,q) 5-branes. The origin of these solutions suggests that they are dual to the Coulomb branch of the $\N=2^*$ theory in the limit when the number of D3 branes becomes infinite and the distance between them is kept fixed. 

However, these solutions are not asymptotically AdS. The dual field theory cannot therefore be interpreted as a UV-finite deformation of the $\N=4$ Super Yang Mills. In a way this theory is similar to the one dual to the Klebanov-Strassler flow \cite{ks}, in that the rank of the gauge group grows as one goes to higher and higher energies. 

This theory has confining, screening, and oblique vacua, much like the one studied by Polchinski and Strassler. In fact, when one of the $\N=1^*$ masses becomes much smaller than the others, the D3 branes polarize into a very elongated ellipsoid \cite{joe}. In the limit when this mass goes to zero while the thickness of the ellipsoid is kept fixed, the ellipsoid degenerates into our cylinder. 

As we will discuss in chapter 2, the background with D4 branes polarized into NS5 branes is dual to a 4+1 dimensional theory with 8 supercharges. Since when the branes are polarized supergravity is valid everywhere, the corresponding phases of the 4+1 dimensional theory have no weakly coupled field theory description. Thus, they can only be described by their supergravity dual, much like the (2,0) and little string theories. In chapter 6 we investigate the phase structure and the objects of this theory. We will find phases in which electric quarks are confined and ``magnetic little strings'' are screened. The exact supergravity dual allows us to find the tension of the confining flux tubes and the masses of the baryons. The theories dual to the nonsupersymmetric exact solutions can also be investigated, and exhibit similar phenomena.

\section{Polarizing D4 branes into NS5 branes - the gauge theory/supergravity picture} 

As explained in \cite{imsy}, supergravity in the near-horizon geometry of a large number $N$ of D4 branes:
\bea
ds^2&=&Z^{-1/2} \eta_{\mu\nu} dx_{\parallel}^2+ Z^{1/2} dx_{\perp}^2 \nn \\
e^{\Phi}&=& g_s Z^{-1/4} \label{g}\\
C_{01234}&=& {1 \over {g_s Z}} \nn 
\eea
describes a certain strongly coupled regime of the field theory living on these branes. When the branes are coincident, $Z={ \pi N g_s \alpha'^{3/2}\over r^3}\equiv {R^3 \over r^3}$.

Both sides of this duality can be perturbed. We can introduce a hypermultiplet mass in the Lagrangian of the field theory; this corresponds in the bulk to turning on a supergravity non-normalizable mode of the RR 2-form and NSNS 3-form field strengths on the directions transverse to the branes \cite{bdhm-bklt}. Indeed, the boundary fermions transform in the {\bf 4} of the $SO(5)$ R symmetry group, and therefore the fermion mass in the {\bf 10} has the same representation as a 2 or 3 form on the 5-dimensional space transverse to the branes.

In this chapter, we perturb the background (\ref{g}) by transverse RR 2 form and NSNS 3 form field strengths, and find the supergravity solution to first order in the perturbation parameter. This solution is the dimensional reduction of the one used in \cite{m5} to explore the polarization of M5 branes into Kaluza Klein Monopoles, so many of the equations will be similar. 

By expanding the IIA supergravity equations of motion: 
\bea
d*F_2=*F_4 \wedge H_3 \nn \nn \\
2d (e^{-2\Phi}* H_3)=F_4 \wedge F_4 -2 d(*F_4\wedge C_1)
\label{b666}
\eea
around the background (\ref{g}), we find that the first order perturbation fields satisfy:
\bea
d\left({1\over Z}(*_5 H_3 + g_s F_2) \right) &=& 0 \nn \\
d\left({1\over Z}(g_s *_5 F_2 + H_3) \right) &=& 0 \label{eom} \\
d F_2 = 0 = d H_3, && \nn 
\eea
where $*_5$ is the flat Hodge operator on the transverse 5-dimensional space\footnote{These equations are very similar to the ones satisfied by the perturbation in \cite{joe} (Eqns. 25,27).}. The metric, dilaton, and 6-form field strength (or its Hodge dual $F_4$) only receive 2'nd order corrections coming from the backreaction of $F_2$ and $H_3$.

We should notice that the form $ {1\over Z}(*_5 H_3 + g_s F_2) $, is harmonic, and thus it is given by its value at infinity. In particular, if one changes $Z$, the form of $F_2$ and $H_3$ might change, but the combination $ {1\over Z}(*_5 H_3 + g_s F_2)$ does not. Also, since $F_4\we F_4=0$, eq. (\ref{b666}) implies that the NSNS 6-form potential is:
\be
d B_6 = e^{-2\phi} *H_3 + *F_4 \we C_1
\ee

We must now relate the precise form of the supergravity perturbations with the fermion bilinears we turn on, by analyzing their $R$-symmetry properties. Luckily, this work has already been done in \cite{m5}. By pairing the 4 worldvolume fermions 
and 4 of the transverse space coordinates into complex combinations
\bea
&&z_1= x^5+i x^6\;\;\;\;\;\; z_2=x^8+i x^9 \label{complex}\\
&&\Lambda_1=\lambda_1+i\lambda_3\;\;\;\;\; \Lambda_2=\lambda_2+i\lambda_4
\eea 
we can see that under an $SO(5)$ rotation $Z^i\rightarrow e^{i\phi^i}Z^i$,
the fermions transform as
\bea
&&\Lambda_1\rightarrow e^{i(\phi^1-\phi^2)/2}\Lambda_1\\
&&\Lambda_2\rightarrow e^{i(\phi^1+\phi^2)/2}\Lambda_2
\eea
Thus, a fermion mass term behaves in the same way under SO(5)
rotations as
\be
T_2= {\rm Re} [m dz_1\we d\bar z_2 + m' dz_1\we dz_2] \equiv {1 \over 2} T_{ij}  dx^i\we dx^j
\label{t2}
\ee

We are interested in  giving mass to half of the worldvolume fermions (together with their corresponding scalars). This preserves $\N=1$ supersymmetry in 4+1 dimension (8 supercharges), and corresponds to $m'=0$. For future reference, we should note that in this case the perturbation breaks the $SO(5)$ R symmetry to $U(1)$.  

Besides $T_2$ there exists another 2-tensor with exactly the same SO(5) transformation properties:
\be
V_2=\frac{1}{2!}(\frac{x^qx^i}{r^2}T_{qj}+ \frac{x^qx^j}{r^2}T_{iq}) dx^i\we dx^j,
\ee
Thus, a general 2 form corresponding to the fermion mass will be a linear combination of $T_2$ and $V_2$, with $r$-dependent coefficients. Similarly, the 3 form will be a combination of the duals of these tensors \footnote{Several useful identities involving these tensors are given in Appendix A.}.
In order to find the 1-form potentials that give the aforementioned 2-form field strength it is also useful to introduce the 1-form: 
\be
S_1 = T_{mn} x^m dx^n 
\ee
satisfying 
\be
d (S_1) = 2 T_2, ~~~~~~d (r^p S_1) = r^p (2 T_2+ p V_2).
\ee

In order to obtain the first order perturbation corresponding to the fermion mass (\ref{t2}) one has to find the form which solves (\ref{eom}) and can be written as a combination of $T_2$ and $V_2$. 
The equations are identical to the ones in \cite{m5}. They have 4 solutions, given in Eq.(2.22) of \cite{m5}. These solutions are the normalizable and non-normalizable modes dual to a fermion mass, and to another irrelevant operator. 

One can see both from the M-theory picture  \cite{m5} or by direct analysis that the non-normalizable mode dual to a fermion mass operator is:
\bea
\nonumber
g_s F_2 &=& Z (2 T_2 - 3 V_2) = d (Z S_1) \nn \\ 
{*}_5 H_3 &=& 3 Z V_2. 
\label{pert1} 
\eea
Note that the actual boundary fermion mass term is not the actual parameter $m$ appearing in this supergravity solution through $T_2$ (\ref{t2}), but is proportional to it through a constant \cite{joe,d2,mariana}. One can use these fields to compute the value of the 6-form NSNS field which couples electrically to NS5 branes:
\bea
d (B_6-C_5\we C_1)&=& e^{-2 \Phi} *H_3 + C_5 \we F_2 = {1\over g_s^2 Z}(*_5 H_3+g_s F_2) \we dx^0 \we dx^1 \we dx^2 \we dx^3 \we dx^4 \nn \\
&=& 2 g_s^{-2} T_2 \we dx^0 \we dx^1 \we dx^2 \we dx^3 \we dx^4
\label{b6}
\eea
Since the expression $B_6-C_5\we C_1$ only depends on the harmonic combination $ {1\over Z}(*_5 H_3 + g_s F_2)$, its value is given by the boundary conditions only, and does not change when $Z$ changes.

To determine whether the solution (\ref{g}),(\ref{pert1}) allows the D4 branes to be polarized into NS5 branes, one must first find the potential of a probe NS5 brane with large D4 charge $n$ (such that $n \ll N$) in the geometry created by the $N$ D4 branes. One can thereafter find the potential for {\bf all} the $N$ D4 branes to be polarized into several NS5 brane shells by treating each shell as a probe in the geometry created by the others. 

The action of type IIA NS5 branes is not an easy one to handle, and was found rather recently \cite{ns5} by reducing the action of the M-theory M5 brane \cite{costin,pst}. Fortunately, the components responsible for the D4 charge have a rather simple form. If all the brane and bulk 3-form fields are turned off, the action becomes:
\bea
S_{BI}&=& \tau_5 \int{d^6\xi e^{-2\Phi}\sqrt{-{\rm det}(g_{ij}-e^{2\phi}\F_i \F_j)}} \label{vbi} \\
S_{WZ}&=& \tau_5 \int{B_6-C_5 \wedge C_1 + C_5 \wedge{\F_1}},\label{vwz}
\eea
where $\F_1 \equiv F_1 +C_1$, and $F_1=d~a$ is the field strength of the scalar living on the NS5 worldvolume. This scalar descends from the M5 brane scalar describing its position on the M-theory circle. Thus, it is no wonder that a nontrivial value of $F^{1}$ corresponds to a nonzero D4 charge. Moreover, we can see from (\ref{vwz}) that to give a circular NS5 brane the D4 charge $n$, one needs to turn on an $F^1$ such that\footnote{The argument for $F^1$ being quantized (as opposed to $\F^1$) is similar to the  one put forth in \cite{wati} for the D-brane worldvolume 2-form.}:
\be
\int_{0}^{2\pi}F_{\phi} d \phi = {n \tau_4 \over \tau_5}
\ee
which implies $F_{\phi} = {n \tau_4 \over 2 \pi \tau_5}=n \alpha'^{1/2} $, where $\tau_4$ and $\tau_5$ are the D4 and NS5 brane tensions respectively. We assume our NS5 brane probe to have D4 charge $n$, and geometry $S^1\times R^5$, where the $S^1$ lies in the $ij$ plane, $i$ and $j$ being two of the transverse directions. The action per unit 4+1 volume in the geometry (\ref{g},\ref{pert1}) has the Born-Infeld part:

\be
V_{BI}= 2 \pi Z^{1/2} \tau_5 g_s^{-2 }\sqrt{g_{\parallel}}\sqrt{g_{\phi\phi} + e^{2 \Phi}\F_{\phi}\F_{\phi}} = 2 \pi Z^{-3/4} \tau_5 g_s^{-2} \sqrt{Z^{1/2} r^2 + g_s^2 Z^{-1/2} \left({{n \tau_4 \over 2 \pi \tau_5} + C_{\phi}}\right)^2}.
\ee

As one can see, the first and the second terms under the square root represent respectively the NS5 and the D4 contribution to the mass of the probe. In the limit we are interested in, the D4 contribution dominates, and thus the Born Infeld action can be Taylor expanded as:
\be
V_{BI} \approx {  Z^{-1} g_s^{-1}  (n \tau_4 + 2 \pi \tau_5 C_{\phi})} +  \frac {2 \pi \tau_5 r^2}{2n g_s^3 \alpha'^{1/2}}
\label{bi}
\ee
The first term represents the gravitational attraction between the $N$ D4 branes sourcing the geometry and the $n$ D4 branes in the probe. The second is the ``left over" mass from the NS5 brane.

The Wess-Zumino action (\ref{vwz}) similarly contains two terms, one representing the RR 3-form mediated repulsion between the D4 branes, and the second coming from the integral of $B_6$ (\ref{b6}) over the worldvolume:
\be
V_{WZ} = - { Z^{-1} (n \tau_4 + 2 \pi \tau_5 C_{\phi}) \over  g_s } - \frac{2 m \pi \tau_5 r^2}{g_s^2}
\label{wz}
\ee

As expected, the leading contributions in the WZ and BI actions coming from interactions between parallel D4 branes cancel each other. Thus, the probe action seems to be given by the 2 remaining terms in (\ref{bi},\ref{wz}):
\be
V_{\rm naive}=\frac {2 \pi \tau_5 r^2}{2n g_s^3 \alpha'^{1/2}}- \frac{2 m \pi \tau_5 r^2}{g_s^2}
\ee

Nevertheless, there exists another term in the action which comes from the interaction of the $n$ D4 branes with the backreaction of the first order fields (\ref{pert1}) on the metric and dilaton. In the next chapter, we will find the exact form of the metric, which allows one to determine this term exactly. However, we can also determine this term using the fact that our setup is supersymmetric, and thus the effective potential for the probe comes from a superpotential. As we will see, the two procedures give the same result, which confirms the validity of our approach.

To obtain the superpotential, it is helpful to express the potential in terms of complex variables. We can also consider a more generic probe, by allowing the transverse circle to degenerate into an ellipse. If $Z_1$ and $Z_2$ (defined as in eq. (\ref{complex})) give the length and orientation of the two semiaxes of the ellipse, then 
$V_{\rm naive}$ becomes
\be
V_{\rm naive}= \frac{\pi \tau_5}{2 n g_s^3 \alpha'^{1/2}}\left(|Z_1|^2+|Z_2|^2-4m n g_s\alpha'^{1/2} {\rm Re} (Z_1 \bar{Z_2})\right)
\ee
and it is not hard to see that it contains two of the three terms coming from the superpotential:
\be
W \sim Z_1 Z_2 - m n g_s \alpha'^{1/2}\frac{(Z_1^2+Z_2^2)}{2}.
\label{W}
\ee
The full potential of the probe is then:
\be
V_n=\frac{\pi \tau_5}{2 n g_s^3 \alpha'^{1/2}}\left(|Z_1-Z_2 mng_s\alpha'^{1/2}|^2+|Z_2-Z_1 mng_s\alpha'^{1/2}|^2\right)
\label{Vn}
\ee
and its minima are at:
\be
Z_1 = mn g_s \alpha'^{1/2} Z_2, \ \ \ \ Z_2 = mng_s \alpha'^{1/2} Z_1
\ee
Evidently the only nontrivial solutions are obtained for:
\be
m=\pm \frac{1}{n g_s \alpha'^{1/2}}
\label{m0}
\ee

This implies that for some special values of the parameter $m$, the radius and orientation of the polarization configuration combine to form a complex modulus. For all other values, the only solution is  $Z_1=Z_2=0$, so there is no polarization.

One should furthermore notice that the polarization potential does not depend on the specific form of the harmonic function $Z$. If the metric is of the form (\ref{g}), the perturbation (\ref{pert1}) is weaker then the background, and the energy of the probe comes predominantly from D4 branes, then $Z$ does not enter the first term of the potential.  Moreover, Eq. (\ref{b6}) implies that $Z$ and does not influence $B_6-C_5\we C_1$,  which gives the second term of the potential. Since the third term is related to the first two by supersymmetry, it likewise has no $Z$ dependence. Thus, the probe potential is independent of the positions of the $N$ D4 branes which source the geometry. Therefore, we can find the full potential of the $N$ D4 branes polarized into several rings of NS5 branes by treating each ring as a probe in the geometry created by the others. The potential is just
\be
V_{\rm full}=\sum_i{V_{n_i}},
\ee
where $n_i$ is the D4 brane charge of the $i$'th tube. For a given $m$, only the tubes with $n_i ={1 \over g_s \alpha'^{1/2} m}$ can have a nonzero radius. It is also possible to superpose several of these tubes, and obtain tubes with $k\times n_i$ D4 branes polarized into $k$ NS5 branes. The energy of such a tube is $k$ times the energy of a simple tube. One can also extrapolate this formula to find that the potential for all $N$ D4 branes to be polarized into one NS5 brane is given by simply replacing $n$ by $N$ in (\ref{Vn}). 

We found a very interesting phenomenon. For certain values of the polarizing field strength the generic configuration consists of several rings of D4 branes polarized into NS5 branes, at generic radii and generic orientations in the 56 and 89 planes. For other values, no solution with polarized branes exists. In the next chapter we will see how this phenomenon beautifully emerges from M-theory.

\section{The exact supergravity solution describing the D4 $\ra$ NS5 polarization}

In this chapter we will find the M-theory description of the polarized
D4 brane configuration found in the previous chapter. This enables us to find the exact
type IIA supergravity solution containing these polarized branes. Moreover, this description provides an intuitive geometric explanation of the moduli space of polarization vacua we found perturbatively.

Let us consider the near horizon 11-dimensional supergravity background of $N$
parallel M5 branes:
\bea
ds^2&=&Z^{-1/3} dx_{\parallel}^2+ Z^{2/3} dx_{\perp}^2 \\ \nonumber
 F^7 &=& d(Z^{-1}) \wedge dx^0 \wedge ... dx^4\wedge dx^{11}.
\label{m1}
\eea
where the branes are aligned along the $0,1,2,3,4$, and $11$ directions, and $Z$ is the harmonic function on the transverse space. When the branes are coincident 
\bea
Z_0= {R_{M5}^3 \over r^3},~~~ r^2= x^i x^i ,~~~ R_{M5}^3= N \pi l_p^3,
\eea
where $i$ runs over the 5 transverse directions. For non-coincident branes, $Z$ is the superposition of the harmonic functions sourced by the individual branes. If the M5 branes are smeared on a circle of radius $r_0$ in the $\rho ~\phi $ plane, $Z$ is given by:
\be
Z = {R_{M5}^3 \over 2 \pi}\int_0^{2\pi}{d\phi \over
{({\bf x}^2 +\rho^2+r_0^2-2 r_0 \rho \cos \phi)^{3/2}}},
\label{smear}
\ee
where ${\bf x}$ denotes the other 3 transverse directions.

As we explained in the Introduction, the polarized state from the previous section can be 
obtained by uniformly distributing the $N$ M5 branes on a transverse circle of radius $r_0$ and performing a dimensional reduction along $\partial_{\tilde{11}} = \partial_{11}+ B \partial_{\phi}$. This is the type of reduction which gives the usual Melvin background.

However, in our case only those twists which identify the upper end of one brane with the lower end of the other are consistent with the setup. These are twists by multiples of $2 \pi \over N$.
The smallest twist joins neighboring M5 branes; the $N$ M5 branes join to form one ``slinky-like'' object, which when reduced to type IIA  becomes a circular NS5 brane with D4 charge $N$. Larger twists join branes which are further apart, and thus give several slinkies. In general, if 
\be
B= {1 \over 2 \pi R^{11}} \left({2 k \pi \over N}\right) 
\label{condB}
\ee
we obtain $N$ D4 branes polarized into $k$ NS5 branes. 

\begin{center}
\begin{figure}[h]
\centerline{\scalebox{.75}{\includegraphics{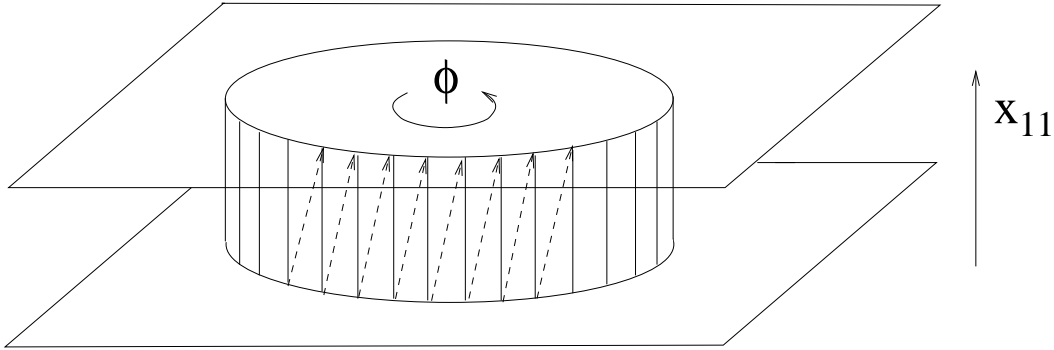}}}
\caption{The twisted compactification of the M5 branes.}
\end{figure}
\end{center}
\vspace{-1cm}



To obtain a Type IIA background, one needs to smear the M5 branes along the circle. Naively, this seems to allow a twist by an arbitrary $B$. Nevertheless, this would give a configuration with a non-integer NS5 brane charge, which is non-physical. The condition that locally the NS5 charge be quantized is equivalent to the constraint (\ref{condB}) on the possible values of the shifts.

For large enough $N$, the discretely arrayed branes are seen in supergravity as smeared. Indeed, if the distance between two M5 branes on the slinky ($ 2\pi r_0 \over N$) is smaller than the radius where the curvature created by one brane becomes larger than the string length, supergravity is only valid away from the slinky. Therefore, the branes appear as effectively smeared. 

The Killing vectors of interest in the 11 dimensional geometry sourced by the smeared branes (\ref{m1}),(\ref{smear}) are $\partial_{11}$,$\partial_{\phi_1}$, and $\partial_{\phi_2}$, where $\phi_1$ is the angular coordinate in the plane of smearing, and $\phi_{2}$ is the angle in an orthogonal plane.

It is possible to obtain a polarized configuration by simply reducing with a twist along $\phi_1$. Nevertheless, such a configuration would not be supersymmetric. To preserve some supersymmetry, we need two twists of equal magnitude.
In the absence of M5 branes, such a reduction would give the supersymmetric flux 5-brane with 8 supercharges found in \cite{perle_si_stromi}. As expected, adding the M5 branes does not spoil the supersymmetry \cite{fs}.

For consistency with the previous chapter, let us choose
the smearing plane to be $x_5 x_8$, and call $\rho_1$ and $\phi_1$ the polar coordinates in this plane. We can also denote by $\rho_2$ and $\phi_2$ the polar coordinates in the orthogonal $x_6 x_9$ plane.

Since the M5 branes are smeared at $\rho_1 =r_0$ in the 58 plane, the harmonic function will only depend on $\rho_1$,$\rho_2$ and $x_7$.
Reducing along the Killing vector $l= \partial_{11}+B_1 \partial_{\phi_1}+B_2\partial_{\phi_2}$ is consistent to performing the 
identifications:
\bea
\nonumber
x^{11} &\sim&  x^{11} + 2 \pi R^{11} n_1 \\  \nonumber
\phi_1 &\sim&  \phi_1 + 2\pi n_2 + 2 \pi n_1 R^{11} B_1 \\ 
\phi_2 &\sim& \phi_2+2\pi n_3 +2 \pi n_1 R^{11} B_2.
\label{id}
\eea 
Supersymmetry requires the $\phi_1$ and $\phi_2$ twists ($B_1$ and $B_2$) to be equal in magnitude \cite{fs,perle_si_stromi}.

The type IIA coordinates descend from 11 dimensional  coordinates with standard periodicity, which are constant along orbits of  the Killing vector $l$:
\be
\tilde \phi_1 =  \phi_1 - B_1  x^{11},~~~~~~ \tilde \phi_2 =  \phi_2 - B_2  x^{11}~,
\ee
By using the relation\footnote{We use Type IIA conventions in which the dilaton is $e^{\phi}$.} between  the M-theory metric and the string frame
metric, dilaton field and the RR 1-form potential:
\bea
ds_{11}^2 = (g_s e^{-\phi})^{2/3} ds_{10}^2+ (g_s e^{-\phi})^{-4/3} (dx^{11}+
g_s C_{\mu}dx^{\mu})^2
\label{redmet}
\eea
we can determine:
\bea
g_s ^{-4/3} e^{4\phi/3}&=& (Z)^{-1/3}+(Z)^{2/3}(\rho_1^2 B_1^2+\rho_2^2 B_2^2) \equiv
\Lambda \\
g_s C_{\tilda \phi_1}&=& \Lambda^{-1} \rho_1^2 B_1 Z^{2/3} \\ \nonumber
g_s C_{\tilda \phi_2}&=& \Lambda^{-1} \rho_2^2 B_2 Z^{2/3} \label{f2ex}\\
\label{metex1}
ds_{10}^2 &=& \Lambda^{1/2}( Z^{-1/3} dx_{\parallel}^2+
Z^{2/3} dx_{\perp}^2)-\Lambda^{-1/2}Z^{4/3}(\rho_1^2 B_1 d \tilda
\phi_1+\rho_2^2 B_2 d \tilda \phi_2)^2,
\label{redus}
\eea
The fields $H_3$ and $F_4$ descend from the  11 dimensional 4-form $\hat F_4$:
\be
\hat F_4= g_s F_4 + dx^{11} \we H_3,
\label{red}
\ee
and are given by:
\bea
g_s F_4&=& *_5 d Z \\
\label{uitelabel}
*_5 H_3&=& (B_1 \rho_1^2 d\tilde \phi_1+ B_2 \rho_2^2 d\tilde \phi_2  )\wedge d Z
\label{*h3}
\eea
where $*_5$ is the flat Hodge dual on the 5-dimensional space transverse to the branes.

It is not hard to obtain from (\ref{f2ex}) and (\ref{*h3}) the first order perturbations found in the previous chapter (\ref{pert1}). The tensors $S_1$ and $V_2$ can be expressed in polar coordinates as:  
\bea
S_1 &=& m (\rho_1^2 d\tilde \phi_1+ \rho_2^2 d\tilde \phi_2) \\
V_2 &=& m (\rho_1^2 d\tilde \phi_1+ \rho_2^2 d\tilde \phi_2) \we (d r /r),
\eea
where $r^2=\rho_1^2+\rho_2^2+x_7^2$. Identifying $m \equiv B_1=B_2$, we can see that to first order in $B$ the exact solution found in this chapter reproduces the one given in (\ref{g},\ref{pert1}). Also, the discrete values of $m$ which allow polarization (\ref{m0}) are the same as the values of $B$ which match the brane ends (\ref{condB}).

\section{The exact supergravity solution describing the F1 $\ra$ D2 polarization}

In this chapter we find the M-theory description of the supersymmetric polarization of $N$ F1 strings into D2 branes. The perturbative analysis of this polarization was performed in \cite{f1}. In that paper it was shown that a large number $N$ of parallel fundamental strings can polarize into cylindrical D2 branes in the presence of transverse RR 2-form and 6-form field strengths \footnote{We use for convenience the conventions of \cite{f1}, $F_6\equiv *\tilda F_4=*(F_4 -C_1 \wedge H_3)$, and $*_8$ is the flat Hodge dual on the transverse space.}:
\bea
g_s  F_2 &=&  Z ( 2 T_2 -6 V_2) \nonumber \\
g_s (*_8 F_6) &=& -Z (6 V_2)
\label{pert2}
\eea
where $T_2$ and $V_2$ are again antisymmetric tensors on the 8-dimensional space transverse to the strings. By grouping the 8 transverse coordinates into 4 complex coordinates:
\be
z^1=x^2+ix^3,~~~z^2=x^4+ix^5,~~~z^3=x^6+i x^7,~~~z^4=x^8+i x^9,
\ee
and by using the SO(8) R-symmetry transformation properties of the fields, it was argued that a perturbation with
\be
T_2= m {\rm Re}(d z^2 d \bar{z}^4)
\ee
preserves 4 supercharges.

The M-theory picture of this polarization is very similar to the one found in the previous chapter. The only change comes from replacing the M5 branes with M2 branes.    

Let us consider the 11 dimensional supergravity background describing the near horizon of a large number $N$ of coincident M2 branes:
\bea
ds_{11}^2 &=& Z_2^{-2/3} d x_{\parallel}^2+ Z_2^{1/3} d x_{\perp}^2  \nonumber \\
\hat F_4 &=& d(Z^{-1})\wedge dx^0\wedge dx^1 \wedge dx^{11} \nonumber \\
Z_0={R_{M2}^6 \over r^6}, &~& r^2=x^i x^i,~~~~~ R_{M2}^6=32 \pi^2 N l_{p}^6
\label{m2}
\eea
where the branes are aligned along $0,1,11$, and $i=2,3,...,9$. 

When the M2 branes are smeared on a circle of radius $r_0$, the only change in the metric above is the harmonic function:
 \be
Z = {R_{M2}^6 \over 2 \pi}\int_0^{2\pi}{d\phi \over
{({\bf x}^2 +\rho^2+r_0^2-2 r_0 \rho \cos \phi)^{6/2}}}= R_{M2}^6{{(\bf x}^2 +\rho^2+r_0^2)^2+2r_0^2 \rho^2 \over({\bf x}^2+(\rho-r_0)^2)^{5/2}({\bf x}^2+(\rho+r_0)^2)^{5/2}}
\label{smear2}
\ee
where ${\bf x}$ denotes the 6 transverse directions perpendicular to the smearing plane.

To obtain the polarized state we again distribute the M2 branes on a circle, and compactify with a twist, as in (\ref{id}). Local D2 charge quantization implies that only certain values of the twist (given by Eq.(\ref{condB}))  give consistent backgrounds. Alternatively, one can see that only twists by multiples of $2\pi \over N$ link an end of an M2 brane with the end of another, like in Figure 1.

We can assume without loss of generality that the M2 branes are distributed in the $x_4x_8$ plane, and introduce polar coordinates $(\rho_1, \phi_1)$ for the $(x_4 x_8)$ plane, and $(\rho_2,\phi_2)$ for the $(x_5 x_9)$ plane. 

If the number of M2 branes is large, supergravity sees them as effectively smeared.
We can therefore dimensionally reduce the background (\ref{m2},\ref{smear2}) along the Killing vector $l= \partial_{11}+B_1 \partial_{\phi_1}+B_2\partial_{\phi_2}$ as in section 3. If $|B_1|=|B_2|$ the resulting background preserves 8 supercharges. 

For completeness, we should note that one can consider a more general reduction, involving twists in 
the $x_2x_6$ and $x_3x_7$ planes as well. For certain values of the twists these reductions can also give supersymmetric backgrounds with polarized branes. The comprehensive analysis done by \cite{fs} for coincident branes applies here without change.

Using the reduction formula (\ref{redmet}), we can determine:
\bea
g_s ^{-4/3}e^{4\phi/3}&=&(Z)^{-2/3}+(Z)^{1/3}(\rho_1^2 B_1^2+\rho_2^2 B_2^2) \equiv
\Lambda \nonumber \\
g_s C_{\tilda \phi_1}&=& \Lambda^{-1} \rho_1^2 B_1 Z^{1/3}  \nonumber\\
g_s C_{\tilda \phi_2}&=& \Lambda^{-1} \rho_2^2 B_2 Z^{1/3}
\label{metex2}\\
ds_{10}^2 &=& \Lambda^{1/2}( Z^{-2/3} dx_{\parallel}^2+
Z^{1/3} dx_{\perp}^2)-\Lambda^{-1/2}Z^{2/3}(\rho_1^2 B_1 d \tilda
\phi_1+\rho_2^2 B_2 d \tilda \phi_2)^2 \nonumber
\eea
Also using (\ref{red}) we obtain:
\bea
F_4 &=&0=\tilde F_4+C_1\wedge H_3 \nonumber \\ 
H_3&=&d(Z^{-1})\wedge dx^0 \wedge d x^1 
\label{exm2}
\eea
Identifying $m\equiv B_1=B_2$ we can verify that to first order in $B$ this exact solution reproduces the perturbative one (\ref{pert2}). 

We should note that the background we found has 8 supercharges, twice the amount found in \cite{f1} by analyzing the 2 dimensional  theory dual to this background. That analysis relied heavily on the study of the free limit of the dual theory, and on assuming the most general type of interactions consistent with the symmetries. Since the Lagrangian is not known, it is possible that it is less general than it was assumed in \cite{f1}, which would explain the increased supersymmetry. One can turn also this observation backwards, and conclude that because its supergravity dual has 8 supercharges, the interactions of the 2 dimensional theory preserve the same amount of supersymmetry. It would be interesting to explore if this is indeed the case, and if it can be seen via a matrix string theory analysis of the type done in \cite{lozano}.

\section{Towards the full Polchinski-Strassler solution}

It is possible to obtain the exact Type IIB solution describing smeared D3 branes polarized into a cylindrical NS5 branes by simply T-dualizing the background (\ref{redus}) along one of the direction parallel to the D4 branes. 

Indeed, the D4 branes become D3 branes smeared along the T-duality direction, while the NS5 branes remain the same. By an $SL(2,Z)$ transformation this configuration can give configurations with D3 branes polarized into (p,q) 5 branes.

These configurations have the same types of fields as in the PS solution. Nevertheless, they have $\N=2$ supersymmetry and have a different topology from the case discussed in \cite{joe}. The NS5 branes we obtain have topology $S^1 \times R^5$, while the ones in \cite{joe} have topology $S^2 \times R^4$. 

To our knowledge there seem to be two major difficulties in obtaining the full PS solution. The first one is finding the exact  $\N=1^*$ supergravity background without the polarized branes, and the second one is finding the modification of this background when the branes are polarized. Our solutions are insensitive to the exact form of $Z$, and seem to suggest that the second step only involves changing the harmonic function $Z$. It would be interesting to see if by applying this intuition to the solution obtained by lifting the 5-dimensional $\N=1^*$ supergravity flow one could find the full PS solution\footnote{A related problem which might be easier to approach would be using an $\N=2$ $AdS_4$ flow \cite{unp} to find the full solution corresponding to M2 branes polarized into M5 branes \cite{m2}.}.

If we choose the T-duality direction $y$ to be $x^4$ the exact solution  (\ref{redus}) becomes:
\bea
e^{\phi} &=& g_s \tilda \Lambda^{1/2} \\
g_s C_2 &=& \tilda \Lambda ^{-1} Z (S_{1}\we d y)\\
\label{uiteasalabel}
*_6 H_3 &=&(S_1 \we d y)\we d Z \\
d^2 s_{10}&=&\tilda \Lambda^{1/2} \left( Z^{-1/2} dx_{\parallel}^2+
Z^{1/2} (dx_{\perp}^2+dy^2)\right)-\tilda \Lambda^{-1/2}Z^{1/2}\left(Z(S_1)^2+(\tilda \Lambda -1) dy^2\right)\\
g_s F_5&=& *_6 d Z ~,
\eea
with
\be
\tilda \Lambda =1+Z(B_1^2 \rho_1^2+B_2^2 \rho_2^2)~~~~{\rm and} ~~~~ S_1 =B_1 \rho_1^2 d\tilda\phi_1+B_2 \rho_2^2 d\tilda\phi_2,
\ee
where the parallel directions are $0123$, $Z$ is given by equation \ref{smear}, and the Hodge dual $*_6$ on the space transverse to the branes has flat indices. 

As a side note we should note that this solution exists even for $B_1 \neq B_2$, when there is no supersymmetry. The exact type IIB solution for a circular D5 brane with large D3 brane charge can be easily obtained using S-duality.

By a chain of T and S dualities the solution describing polarized F1 strings can also be brought to describe an $AdS_5 \times S^5$ geometry perturbed with metric and 5-form components which break the 3+1 dimensional Lorentz invariance of the boundary theory. These perturbations preserve 8 supercharges, and allow the D3 branes to develop dipole moments corresponding to D3 brane charge along directions transverse to the branes. 

\section{ More about the theory on the D4 branes}

As we explained in the previous chapters, the strongly coupled theory
dual to the supergravity background with polarized branes is related to
the 4+1 Super Yang Mills theory living on the D4 branes. As it is well
known, this theory is not renormalizable, and becomes strongly coupled in
the UV. In that regime it can be described by string theory on the background
(\ref{metex1}), which can be thought of as the dual of the UV completion of this theory.

By turning on the supergravity modes corresponding to fermion masses, the
UV completion is modified, and can in some cases include polarized branes.
In these cases, the supergravity solution valid everywhere, and thus
there is no regime where the boundary theory is weakly coupled. 
When there are no polarized branes, the supergravity background becomes
again singular, and the IR of the field theory becomes weakly coupled.

For fermion masses allowing brane polarization ($m \sim {k \over Ng_s \alpha'^{1/2}}$) one can pass from a phase where the theory has a weakly coupled field theory description (as a mass-deformed $\N=1$ Super Yang Mills theory in 4+1 dimensions) to a phase where there is no weakly coupled field theory description, by simply changing the polarization radius.


Note that such theories are not new. They are similar to the mysterious ``little string theories'', or 
to the (2,0) theory, which can also be defined only via
their weakly coupled supergravity duals. In the case of these theories however, one can
also study them via DLCQ; it would be interesting to see if aspects of the 4+1
dimensional theories discussed here are amenable to a similar discussion.

The purpose of this chapter is to learn as much as possible about these
theories by studying their supergravity duals. The first thing to notice
is that these theories have 8 supercharges. One can see this both directly (a mass to a chiral superfield in 4+1 dimensions preserves $\N=1$ supersymmetry) or by noticing that the exact supergravity dual of these theories has 8 supercharges \cite{fs}.

{\bf Quarks and Little Strings}

As in the case of D3 branes, the ends of objects ending on the branes
are  ``states'' in the boundary theory. As both F1 strings and D2 branes can
end on a D4 brane, this theory will have both ``quarks" and ``little strings".

Thus, an infinite F1 string ending on a D4 brane can be interpreted as a
quark. In the confining phase, the energy of the flux tube between two
such quarks is given by the energy of an F-string with its ends on the boundary,
lowered into the bulk \cite{neaion}. Since the bulk contains only NS5
branes, the F string can never attach to them, and thus the string
energy is proportional to the separation of its ends. This indicates that 
quarks are confined.

One can also see that the D2 brane ``little strings", are
screened. The generalized Wilson surface which describes the properties 
of these strings is given by the energy of a D2
brane lowered in the bulk \cite{neaion,m5}. Since this D2 brane can
attach itself to the NS5 brane, there is no energy cost to move the two
``little strings" apart. Therefore the little strings are screened. Since we are
in a phase where the quarks are confined, it is appropriate to call the
little strings ``magnetic little strings".

We should note that if we were in a phase with the D4 branes polarized
into a D6 brane (such a state can only be obtained without supersymmetry \cite{m5}), 
the strings would be confined and the quarks would be screened. These two phases
are very reminiscent of the ones in 3+1 dimensional theories.
Nevertheless, if we insist on preserving 8 supercharges only the
``magnetic" phase is present. 

It is quite easy to find the tension of the confining flux tube. When the
quarks are far apart, the bulk string joining them is composed of
essentially  two vertical segments, and one segment sitting near the
polarized branes. The energy of the two vertical segments is essentially
constant, and therefore the flux tube tension is given by the tension of
an F1 string sitting near the shell.

It is possible to extract the components of the near-shell geometry from the exact solution. At $\rho_1=r_0+\epsilon$, the harmonic form (\ref{smear}) becomes:
\be
Z_{\rm near ~shell}= \frac{R^3_{M5}}{ \pi r_0 \epsilon ^2},
\ee 
and therefore $g_{\parallel}= r_0 B + O(\epsilon)$, and $e^{-\phi} \sim \epsilon$.

Thus, the flux tube tension is: 
\be
T_{\rm flux ~tube} = \sqrt{-g_{00}g_{11}}~|_{\rm near ~shell} = r_0 B,
\ee
independent of the 't Hooft coupling of the boundary theory. Note that as $r_0 \rightarrow 0$ the weakly coupled infrared region is recovered, there is no confinement, and the string tension becomes zero as expected.


One can also see that the magnetic little strings are
screened, by estimating the energy of a D2 brane in the near shell limit:
\be
V_{\rm D2} = e^{-\Phi} \sqrt{-g_{00}g_{11}g_{22}}~|_{\rm near~shell}\sim \epsilon \rightarrow 0
\ee
This confirms the string theory intuition outlined above.

{\bf Baryons}

Since the weakly coupled theory has an $SU(N)$ gauge
symmetry, one expects that in the confining phase a baryon made of $N$ quarks 
is a free object.
One can easily see that bulk dual of a baryon in the unperturbed field theory is a D4 brane wrapping the warped
4-sphere transverse to the D4 branes. Nevertheless, unlike its 3+1 
dimensional ``cousin" \cite{w}, this baryon is not
stable because of the lack of conformal invariance. It tends to
slide off towards the infrared and self-annihilate.

Nevertheless, when the D4 branes are polarized, the D4 brane baryon 
sliding towards the infrared crosses
the polarized configuration at a finite radius. Via the Hanany-Witten
effect, the resulting baryon is a D2 brane ending on the NS5-D4 shell, and
filling the 2-ball whose boundary is the polarization circle. It is not
hard to see that $N$ fundamental strings can end on the junction between
the D2 brane and the NS5-D4 shell.

Indeed, by investigating the NS5 brane action \cite{ns5} (formulas 54,55), we can
see that the D2 brane ends source a nonzero NS5 worldvolume 3-form $db_2$, and the dissolved D4 branes create a nonzero worldvolume 1-form $\F_1$. 
The anomaly given by the term
\be
db_2 \we \F \we B_2
\ee
under the gauge transformation $\delta B_2= d \chi_1$ is proportional to the number of dissolved D4 branes ($N$), and 
can only cancel if $N$ F1 strings end on the NS5-D2 junction. 
Therefore, the D2 brane filling the 2-ball inside the polarization circle is indeed the baryon of this theory.

One can also estimate the dependence of the mass of this baryon on the parameters of the theory. Assuming the order of magnitude of $Z$ to be $R^3/r_0^3$, we find the tension of the D2 brane to be:
\be
M_{\rm baryon}= \tau_{D2} \int_0^{r_0} d \rho_1 d \phi_1 e^{-\Phi} \sqrt{-g_{00}g_{\rho_1 \rho_1} g_{\phi_1 \phi_1}} \approx \sqrt{N^3 r_0 g_s}.
\ee

{\bf Domain Walls} 

As we have seen, our theory has a moduli space of polarization vacua, and one would expect two different vacua to be separated by a domain wall. Nevertheless, these domain walls do not exist. The easiest way to see this is to remember that the tension of a domain wall is given by the difference of the superpotentials in the two vacua it separates. Nevertheless,  since the superpotentials are zero in all vacua (\ref{W}), the domain wall tension is zero.

{\bf Condensates}

One can find the value of normalizable modes of supergravity fields by simply expanding the exact solution for large $r$. These normalizable modes correspond to vacuum expectation values of certain operators in the boundary theory \cite{bdhm-bklt}. Knowing the bulk-boundary dictionary allows one to find all these expectation values with minimal effort.

\section{Conclusions}

We have investigated the polarization of D4 branes into NS5 branes both by perturbing their near-horizon geometry and performing a Polchinski-Strassler type analysis, and by investigating the M-theory origin of this polarization.

This enabled us to obtain the exact supergravity solutions describing this polarization, to our knowledge the first exact solution which contains polarized branes and has a field theory dual. We also obtained the exact solution describing the polarization of F1 strings into D2 branes. A generic such solution contains several concentric polarized configurations, of arbitrary radii, and arbitrary orientations.

We then used T-duality to obtain type IIB solutions with 8 supercharges describing smeared D3 branes polarized into concentric cylindrical (p,q)  5 branes. By a chain of T and S dualities it is also possible to obtain asymptotically AdS solutions where the D3 branes develop a transverse D3 dipole moment. These solutions are the first exact IIB supersymmetric solutions with polarized branes and a field theory dual. 

In the last chapter we investigated some of the properties of the supersymmetric 4+1 dimensional theory dual to the D4-NS5 exact background, and gave string theory descriptions of the objects this theory contains: quarks, magnetic little strings, baryons, domain walls, etc.

The solutions found in this paper belong to the same universality class as the exact Polchinski-Strassler solution, and we hope that the ideas presented here will be useful steps towards finding this solution.

{\bf Acknowledgments} 
We would like to thank Joe Polchinski, John Schwarz, Per Kraus, Roberto Emparan, Nick Warner, Anton Kapustin, and Radu Roiban for useful discussions and suggestions. The work of I.B. was supported in part by the the NSF Grant PHY00-99590. The work of C.C. was supported in part by the DOE Grant DE-FG03-92-ER40701.

\newpage
\section{Appendix A - Tensor Spherical Harmonics}
We give several useful relations between the transverse space tensors used in this paper. If the transverse space is 5-dimensional, and we are interested in describing antisymmetric 2-form and 3-form harmonics, they depend on the tensors

\bea
T_2 &= & {1 \over 2 !} T_{mn} dx^m \we dx^n \label{T2}\\
V_2 & =& \frac{1}{2!}(\frac{x^qx^i}{r^2}T_{qj}+ \frac{x^qx^j}{r^2}T_{iq}) dx^i\we dx^j \label{V2}\\
T_3&=&*_5 T_2  \\
V_3&=&\frac{1}{3!}(\frac{x^q x^m}{r^2} T_{qnp}+{\rm{2\; more}}) dx^m\we dx^n \we dx^p 
\eea
which satisfy:
\bea
T_2-V_2&=&*_5 V_3\\
d(\ln r)\we V_3&=&0\\
d(\ln r)\we *_5 V_3&=&d(\ln r)\we *_5 T_3\\
d(V_3)&=&-3d(\ln r)\we T_3\\
d(*_5 V_3)&=&2 \ln r\we *_5 T_3
\eea
In order to express the 1-form potential it is also useful to introduce the transverse space 1-form:
\be
S_1 = T_{mn} x^m dx^n \\
\label{s1}
\ee
satisfying
\bea
d (S_1) &=& 2 T_2\\
d (r^p S_1) &=& r^p (2 T_2+ p V_2)
\label{s11}
\eea

If the transverse space is 8 dimensional one can similarly introduce 2-form and 6-form tensors. \footnote{For the precise formulas see the Appendix in \cite{f1}.} We give all the fields in terms of $T_2,V_2$ and $S_1$, and equations (\ref{T2},\ref{V2},\ref{s1},\ref{s11}) are the only ones needed.

\section{Appendix B - Consistency Checks}  

As explained in \cite{imsy} in the case of a large $N$ number of D4 branes there exists a decoupling limit, $\alpha' \to 0$, keeping $g_s (\alpha')^{1/2}$  fixed, where the field theory on the branes decouples from the theory in the bulk. 

The type  II A supergravity solution
can be trusted  \cite{imsy} in the region:$~~~$$\frac{\alpha'}{N} \ll r \ll N^{1/3} \alpha'$. For smaller $r$ the curvature becomes too large, and the weakly coupled description of the physics is provided by the 4+1 dimensional Super Yang Mills theory. For larger $r$ the dilaton becomes too large, and weakly coupled description of the physics is provided by 11 dimensional supergravity.

The condition for the validity of the perturbative calculations done in the first 2 sections is: $ \frac{|F_2|^2}{|F_6|^2} \ll 1$:
\bea
|F_2|^2&=& F_{ij}F_{ij} g^{ii}g^{jj} \sim g_s^{-2} m^2 Z \sim \frac{m^2 N g_s^{-1} \alpha'^{3/2}}{r^3}\\ \nonumber
|F_6|^2&=&F_{01234r}F_{01234r}g^{00}...g^{44}g^{rr} \sim \frac{1}{r^2} \nonumber
\eea
Thus the perturbation is small if $m^2 N g_s^{-1} \alpha'^{3/2} \ll r$. For the smallest mass which allows for a moduli space this is equivalent(in the decoupling limit) to $ \frac{\alpha'}{g_s^2N} \ll r$, which is trivially satisfied.  For the other masses which allow polarization the perturbative calculation are valid for $ \frac{k^2 \alpha'}{g_s^2N} \ll r$. 

Finally let us consider the regime where the M5 branes become effectively smeared. The curvature near a single M5 brane is large in string units for distances of order $ \alpha'$. Therefore the smearing approximation is justified for $\frac{2\pi r_0}{N} \ll \alpha'$ . This constraint is  satisfied in the energy region of interest.

The regime where the M5 branes are seen as smeared is the same as the regime where the D4 contribution to the energy of the polarized configuration is dominant (\ref{bi}). Outside this region, both the supergravity perturbative approach in chapter 2 and of the exact solution in chapter 3 stop being valid. Nevertheless, it is quite likely that a solution with polarized branes still exists. Indeed, the radius and orientation of these solutions parameterize the moduli space of a 4+1 dimensional theory with 8 supercharges. It is quite unlikely that by taking the branes further away the moduli space would go away. It would be interesting to investigate if this is indeed the case, and to see what the  D4-NS5 soliton becomes in this regime.

\end{document}